\documentclass[utf8]{frontiersFPHY} 

\usepackage{graphicx} 
\usepackage{dcolumn}
\usepackage{bm}
\usepackage{amssymb}
\usepackage{amsmath}
\usepackage{wasysym}
\usepackage{gensymb}
\usepackage{color}
\usepackage{hyperref}
\usepackage[toc,page]{appendix}
\usepackage{float}
\usepackage{flushend}
\usepackage{soul}
\usepackage{bibentry}
\usepackage{url,hyperref,microtype,subcaption}
\usepackage[onehalfspacing]{setspace}
\hypersetup{
    colorlinks,%
    citecolor=blue,%
    filecolor=blue,%
    linkcolor=blue,%
    urlcolor=blue
}

\def\keyFont{\fontsize{8}{11}\helveticabold }
\def\firstAuthorLast{Kumar {et~al.}} 
\def\Authors{Rohit Kumar\,$^{1,*}$, Laur{\`e}ne Jouve\,$^{1}$, Rui F. Pinto\,$^{1}$ and Alexis P. Rouillard\,$^{1}$}
\def\Address{$^{1}$Institut de Recherche en Astrophysique et Plan{\'e}tologie, Universit{\'e} de Toulouse, CNRS, UPS, CNES, 31400 Toulouse, France}
\def\corrAuthor{Rohit Kumar}
\def\corrEmail{rohitkumar.iitk@gmail.com}

\begin{document}
\onecolumn
\firstpage{1}

\title[Production of sunspots and their effects on the corona and solar wind]{Production of sunspots and their effects on the corona and solar wind: Insights from a new 3D flux-transport dynamo model}

\author[\firstAuthorLast ]{\Authors} 
\address{} 
\correspondance{} 
\extraAuth{}

\maketitle

\begin{abstract}
\section{}
We present a three-dimensional numerical model for the generation and evolution of the magnetic field in the solar convection zone, in which sunspots are produced and contribute to the cyclic reversal of the large-scale magnetic field. We then assess the impact of this dynamo-generated field on the  structure of the solar corona and solar wind. This model solves the induction equation in which the velocity field is prescribed. This velocity field is a combination of a solar-like differential rotation and meridional circulation. We develop an algorithm that enables the magnetic flux produced in the interior to be buoyantly transported towards the surface to produce bipolar spots. We find that those tilted bipolar magnetic regions contain a sufficient amount of flux to periodically reverse the polar magnetic field and sustain dynamo action. We then track the evolution of these magnetic features at the surface during a few consecutive magnetic cycles and analyze their effects on the topology of the corona and on properties of the solar wind (distribution of streamers and coronal holes, and of slow and fast wind streams) in connection with current observations of the Sun.

\tiny
 \keyFont{ \section{Keywords:} solar magnetic cycle, mean field dynamo, flux-transport, sunspots, field reversal, solar corona} 
\end{abstract}

\section{Introduction}

The Sun's magnetism is responsible, among other things, for the production of sunspots and polar field reversals~\citep{Ossendrijver:AAR2003}; which result in coronal heating and solar wind. This activity is connected to what is now currently called space weather and affects the Earth's environment, in particular satellite operations and telecommunications.   

Among the different dynamo models applied to the Sun's magnetic field, one has been particularly successful at reproducing many of the observed solar features: the Babcock-Leighton (BL) dynamo model, proposed in the 60's by Babcock and Leighton~\cite{Babcock:APJ1961,Leighton:APJ1969}. In this model, the toroidal field is generated near the base of the solar convection zone (SCZ) by the shearing induced by the differential rotation. This toroidal field then gets transported, due to the magnetic buoyancy, to the solar surface to produce bipolar magnetic regions (BMRs or sunspots)~\citep{Charbonneau:LRSP2005}. Because of the Coriolis force acting on the rising toroidal structures, the emerged sunspot pairs possess a tilt with respect to the East-West direction such that the leading spot is located at a lower latitude than the trailing one~\citep{D'Silva:AA1993}. These tilted BMRs are then advected by the surface flows and are subject to the turbulent diffusion induced by small-scale convective motions. If some trans-equatorial cancellation is allowed for the leading spots of both hemispheres, the net flux advected towards the poles is then of one particular polarity, opposite to the one of the polar field. This new flux will thus reverse the polar field (this is precisely the BL mechanism) which in turn will produce the toroidal field of the next cycle. If the generation of the toroidal field through differential rotation is well accepted, the crucial role of spots to reverse the poloidal field is more debated. However, a recent study by Dasi-Espuig {et al.}~\cite{Dasi-Espuig:AA2010} in 2010 has shown that a correlation indeed exists between an observable measurement of the Babcock-Leighton mechanism and the strength of the next solar cycle. This would indicate that the sunspots indeed play an important part in the dynamo cycle.

The solar dynamo has been studied for decades using two different approaches: either two-dimensional (2D) kinematic mean-field dynamo models where the evolution of the large-scale magnetic field is computed for a prescribed velocity field~\citep{Moffatt:book,Krause:book}, or three-dimensional (3D) global models where the evolution of both the velocity and the magnetic fields are studied by solving the full set of magnetohydrodynamic (MHD) equations in which the velocity and magnetic fields interact nonlinearly (see reviews by Miesch and Toomre ~\cite{Miesch:ARFM2009} and Brun et al.~\cite{Brun:SSR2015}). Several 2D BL flux-transport dynamo models have been used to study the solar magnetism~\citep{Dikpati:APJ1999,Jouve:AA2007}. But, in those 2D models, it is assumed that the BL process is the source for the poloidal field and thus a source term is added to the poloidal field equation in an ad hoc way. In 3D models, the strong toroidal structures built in rapidly-rotating stars can become buoyant~\cite{Nelson:APJ2013,Fan:APJ2014} but rarely rise all the way to the top of the computational domain and, consequently, those models do not produce spots. It is thus not possible in such ``spotless" models to assess the potential role of spots in the large-scale field reversals and to study the impact of such evolving spots on the coronal structure. 

We propose here to develop a 3D kinematic solar dynamo model in which the toroidal field owes its origin to the shearing of the poloidal field by the differential rotation and where an additional buoyancy algorithm is implemented to force strong toroidal regions to rise through the convection zone and produce spots. The first objective here is to see the BL mechanism at play, i.e., study the ability of these tilted BMRs to reverse the polar field and serve as a seed for the next toroidal field. In this first step, we will not try to particularly calibrate our model on real solar observations but this model, strongly relying on surface features like sunspots, will serve as a first step towards introducing data assimilation for long-term forecasting. Similar models have been developed recently, using two approaches which differ in the way BMRs are produced at the surface. Miesch and Dikpati~\cite{Miesch:APJL2014}, followed by Miesch and Teweldebirhan~\cite{Miesch:ASR2016}, used a double-ring algorithm, putting BMRs directly at the solar surface and extracting an equivalent toroidal flux at the base of the convection zone. On the other hand, Yeates and Mu{\~n}oz-Jaramillo~\cite{Yeates:MNRAS2013} used a more realistic method in which an additional velocity field is responsible for the transport of toroidal flux from the base of convection zone to the solar surface. Miesch and Teweldebirhan~\cite{Miesch:ASR2016} indeed obtained cyclic reversals of the magnetic field using their ``spot-maker" algorithm where the step of flux rising from the bottom of the convection zone is not explicitly modeled. In the more realistic model of Yeates and Mu{\~n}oz-Jaramillo~\cite{Yeates:MNRAS2013}, only one solar cycle was modeled and the possibility to have a self-sustained dynamo was not studied. 

We here describe a model similar to Yeates and Mu{\~n}oz-Jaramillo~~\cite{Yeates:MNRAS2013} but where several cycles are computed. We then reconstruct the evolution of the coronal magnetic field in response to the modeled dynamo field and study how coronal holes vary in size and position and how the open flux at various latitudes evolve along the magnetic cycle. Those quantities are particularly important to assess the 3D structure of the solar wind and how the wind speed varies during a full magnetic cycle. We note that we are not trying here to simulate the whole complexity of the solar magnetic field evolution from the base of the convection zone to the corona and solar wind structure, we are instead showing a proof-of-concept that important large-scale features can be reproduced by our simplified models and that, thanks to those simplifications and the small computational cost, data assimilation could be easily introduced in this integrated model to help forecasting the future evolution of the large-scale solar magnetic field.

The rest of the paper is organized as follows: we describe the details of our new 3D kinematic dynamo model in Sec.~\ref{sec:num_detail}. In Sec.~\ref{sec:results}, we present the simulation of the self-sustained Babcock-Leighton dynamo. In Sec.~\ref{sec:corona_wind}, we focus on the coronal magnetic topology and the solar wind during a full solar cycle. Finally, in Sec.~\ref{sec:conclusion}, we summarize our results.

\section{The 3D kinematic dynamo model}
\label{sec:num_detail} 

\subsection{The numerical code}
\label{sec:num_code}

We solve the magnetic induction equation in a 3D spherical shell. This equation reads:
\begin{eqnarray}
\frac{\partial \mathbf{B}}{\partial t} = \nabla \times (\mathbf{v} \times \mathbf{B}) - \nabla \times (\eta \nabla \times \mathbf{B}),
\end{eqnarray}
where $\mathbf{B}$ is the magnetic field, $\mathbf{v}$ is the prescribed velocity field, and $\eta$ is the magnetic diffusivity. In our simulations, the prescribed velocity field is a combined effect of the differential rotation and the meridional circulation.   

We perform dynamo simulations in a spherical shell geometry using the pseudo-spectral solver MagIC~\citep{Wicht:PEPI2002,Gastine:Icarus2012}. MagIC employs a spherical harmonic decomposition in the azimuthal and latitudinal directions, and Chebyshev polynomials in the radial direction. For time-stepping, it uses a semi-implicit Crank-Nicolson scheme for the linear terms and the Adams-Bashforth method for the nonlinear terms.

The inner and outer radii of the computational shell are $[0.65, 1.0] \, R_{\odot}$, where $R_{\odot}$ is the solar radius. We choose $N_r=64$ grid-points in the radial, $N_\theta=128$ points in the latitudinal, and $N_\phi=256$ points in the longitudinal direction. Simulations are performed by considering insulating inner and outer boundaries for the spherical shell. We employ an initial magnetic field which is the combination of a strong toroidal and a relatively weak poloidal magnetic field. The strong toroidal field is chosen so that the magnetic flux due to the BMRs generated at the surface would be sufficient to reverse the polar field of the first cycle. We thus make sure that the initial conditions are favorable to produce a first reversal. We however note that the steady-state solutions are not very sensitive to our choice of initial conditions.

\subsection{Velocity field and diffusivity profile}
\label{sec:fl_tr_model}  

In this section, we present the ingredients of the dynamo model we used, in particular the velocity field and the magnetic diffusivity profile. 

We prescribe the velocity field such that the flow has a differential rotation similar to that observed in the Sun through helioseismology~\citep{Schou:APJ1998}. In addition, the velocity field also consists of a meridional flow, which is poleward near the surface and equatorward near the base of the convection zone. The meridional flow is an important ingredient since it is thought to be responsible for the advection of the effective magnetic flux of the trailing spots towards the poles to reverse the polarities of the polar field~\citep{Wang:APJ1989,Dikpati:AA1994,Dikpati:SP1995,Choudhuri:SP1999}.  
Both the expressions of the differential rotation and the meridional flow are slightly modified versions of the ones used in~\citep{Jouve:AA2008}.

\begin{figure}[htbp]
\centering
\includegraphics[scale=1.0]{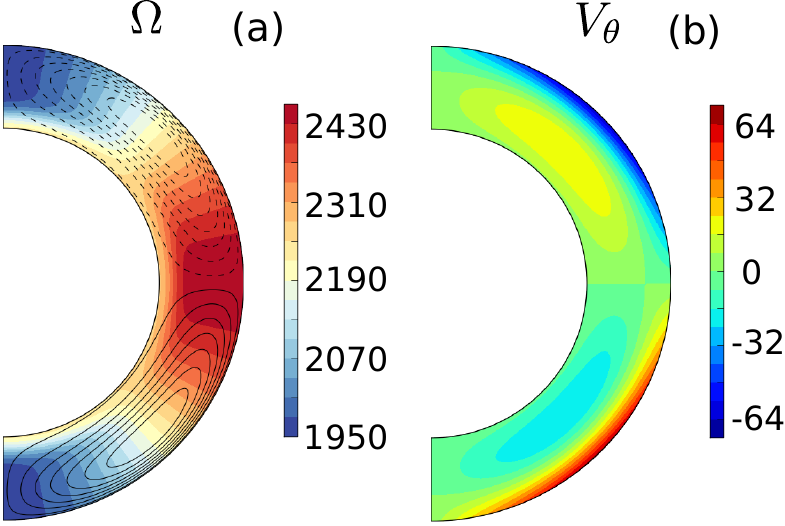}
\caption{Plots of longitudinally averaged (a) differential rotation ($\Omega$) superimposed with the meridional flow (dashed lines: counter-clockwise, solid lines: clockwise) and (b) latitudinal velocity ($V_\theta$) in the convection zone. Here $\Omega$ and $V_\theta$ are in non-dimensional units. In dimensional units, $V_\phi (=\,r \, sin \, \theta \; \Omega) \approx 2 \; km/s $ at the surface near the equator and $V_\theta \approx 20 \; m/s$ at the surface (latitude: $\pm 45 \degree$).}
\label{fig:diff_mer_fl_num}
\end{figure}

In Fig.~\ref{fig:diff_mer_fl_num}(a), we show the differential rotation superimposed with the meridional flow in our simulations. The rotation is strongest in the equatorial region, and it decreases as we move towards the poles. The observation of solar velocities suggest that the magnitude of the averaged meridional circulation at the solar surface ($\approx 20 \; m/s$) is almost $100$ times smaller than the rotational velocity at the equator ($\approx 2 \; km/s$)~\citep{Roudier:AA2012}. Assuming those observed values and choosing the solar radius as the characteristic length-scale, the Reynolds numbers associated with the longitudinal velocity ($V_\phi= \,r \, sin \, \theta \; \Omega$) and with the latitudinal velocity ($V_\theta$) are $\mathrm{Re=V_\phi R_\odot/\eta_s} \approx 7000$ (maximum $V_\phi$ at the surface near the equator) and $\mathrm{Re_p=V_\theta R_\odot/\eta_s} \approx 70$ (maximum $V_\theta$ at the surface), respectively, where $\eta_s$ is the value of the magnetic diffusivity at the surface. The rotation rate and the latitudinal velocity are illustrated in Fig.~\ref{fig:diff_mer_fl_num}.  

The magnetic diffusivity is a two-step function similar to the one defined in Yeates and Mu{\~n}oz-Jaramillo~\citep{Yeates:MNRAS2013}. The only differences lie in the values chosen for the various diffusivities. The value of the diffusivity is smallest near the stable radiative zone $\eta_c= 5 \times 10^{10} cm^2.s^{-1}$, larger in the bulk of the turbulent convection zone $\eta_0= 10^{12} cm^2.s^{-1}$ and the largest near the outer surface $\eta_s= 2 \times 10^{12}cm^2.s^{-1}$, where the turbulent diffusion induced by small-scale convective motions is thought to be even more enhanced.

We now describe how the rise of toroidal flux through magnetic buoyancy was implemented in the code. 

\subsection{Magnetic buoyancy algorithm}
\label{sec:buoy_algo}
 
Following Yeates and Mu{\~n}oz-Jaramillo~\cite{Yeates:MNRAS2013}, we employ an additional velocity field which transports the magnetic flux produced in the solar interior towards the surface. This velocity models the effects of magnetic buoyancy  which is responsible for the rise of strong magnetic field from the base of the convection zone to the surface. This additional velocity contains two components: a radial one to model the rise and a vortical one to produce a tilt. The radial velocity as a function of longitude and latitude (${\phi}, {\theta}$) is described as   
\begin{equation}
V_r = V_{r0} \; exp{\left[-\left\lbrace \left(\frac{\phi - {\bar \phi}}{\sigma_\phi}\right)^2 + \left(\frac{\theta - {\bar \theta}}{\sigma_\theta}\right)^2 \right\rbrace \right]},
\end{equation}   
where (${\bar \phi}, {\bar \theta}$) corresponds to the apex of the rising flux tubes, $V_{r0}$ is the amplitude of the velocity (the corresponding Reynolds number is $500$), $\sigma_\theta = \sigma_\phi =5$ degrees. In our model, ${\bar \theta}$ and ${\bar \phi}$ are randomly chosen, such that the positions of the emerging regions at the surface are random. We choose in our particular model to have emergence of 32 BMRs every 4.5 months. Taking into consideration the latitudes of the observed sunspots, we choose the values of ${\bar \theta}$ to stay between $[-35, +35]$ degrees. 

For representing the effect of the Coriolis force due to the Sun's rotation, we then incorporate an additional vortical velocity, which imparts a tilt to the emerging magnetic flux ropes. The vortical velocity is a combination of latitudinal and longitudinal velocity components. The vortical velocity is tuned so that the tilt in the BMRs is clockwise in the northern hemisphere, and anti-clockwise in the southern hemisphere and corresponds to values observed at the solar surface (between $4 \degree$ and $14 \degree$~\cite{Wang:SP1989}). The vortical velocity is also designed such that the tilt angle increases as we move towards the poles, following Joy's law~\citep{Hale:APJ1919}. We do not give the expression of this vortical component as it is identical to the one described by Yeates and Mu{\~n}oz-Jaramillo~~\cite{Yeates:MNRAS2013}.

In agreement with the physics of magnetic buoyancy instabilities, the additional velocity field is applied only for a toroidal magnetic field $B_\phi > B_{\phi}^l$. In dimensional units, $B_\phi^l \approx 4\times10^4$ Gauss. Below this value, the field is thought to be strongly influenced by the Coriolis force and thus rise parallel to the rotation axis \cite{Choudhuri:APJ1987}. At the other end, we also suppress the effects of the additional velocity field for $B_\phi > B_{\phi}^h$. In dimensional units, $B_\phi^h \approx 1.4 \times 10^5$ Gauss. Indeed, when the field is too strong, it is assumed to rise very rapidly to the surface without being affected by the Coriolis force. As a consequence, the produced BMRs will not be tilted and will thus not take part in the reversal of the polar field since the longitudinally averaged net flux will be zero \cite{D'Silva:AA1993}.

\section{Dynamo solution: production of BMRs and polarity reversals}
\label{sec:results}

\subsection{A self-sustained cyclic dynamo}

\begin{figure}[htbp]
\centering
\includegraphics[scale=1.0]{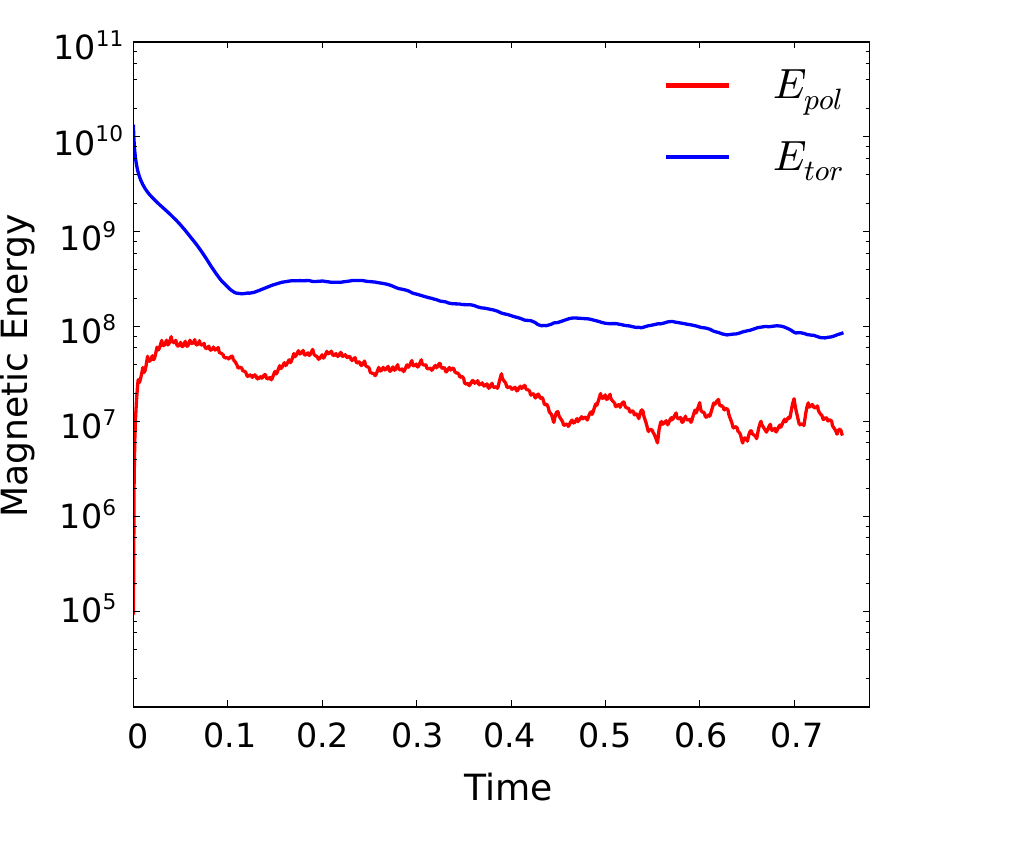}
\caption{The non-dimensional poloidal and toroidal magnetic energy with time (magnetic diffusion). Time $t=1$ corresponds to one magnetic diffusion time (using the surface value for the magnetic diffusivity), which is approximately $77$ years.}
\label{fig:mag_en_32}
\end{figure}

Starting with the initial conditions described in Sec.~\ref{sec:num_code}, we follow the time evolution of the magnetic field components in our new 3D kinematic dynamo model.

In Fig.~\ref{fig:mag_en_32}, we show the time-evolution of the toroidal and poloidal components of the magnetic energy. We clearly see here an exponential growth of the poloidal magnetic field and then a saturation of both components, as expected in a self-sustained dynamo. As discussed in Sec.~\ref{sec:buoy_algo}, magnetic buoyancy acts on the toroidal field only when $B_{\phi}^l < B_\phi < B_{\phi}^h$, i.e., there is upper and lower cutoffs on the emerging toroidal flux. These cutoffs introduce an additional nonlinearity in the system, which causes the saturation of the magnetic energy with time. This nonlinearity plays a similar role as the quenching term in 2D kinematic dynamo models. The saturation level of the  magnetic energy depends on both the upper and the lower cutoffs on $B_\phi$. Indeed, when more flux is allowed to reach the surface (i.e., when $B_{\phi}^l$ is lower and $B_{\phi}^h$ is higher), the magnetic energy naturally saturates at a higher level.

\begin{figure*}[htbp]
\centering
\includegraphics[scale=0.45]{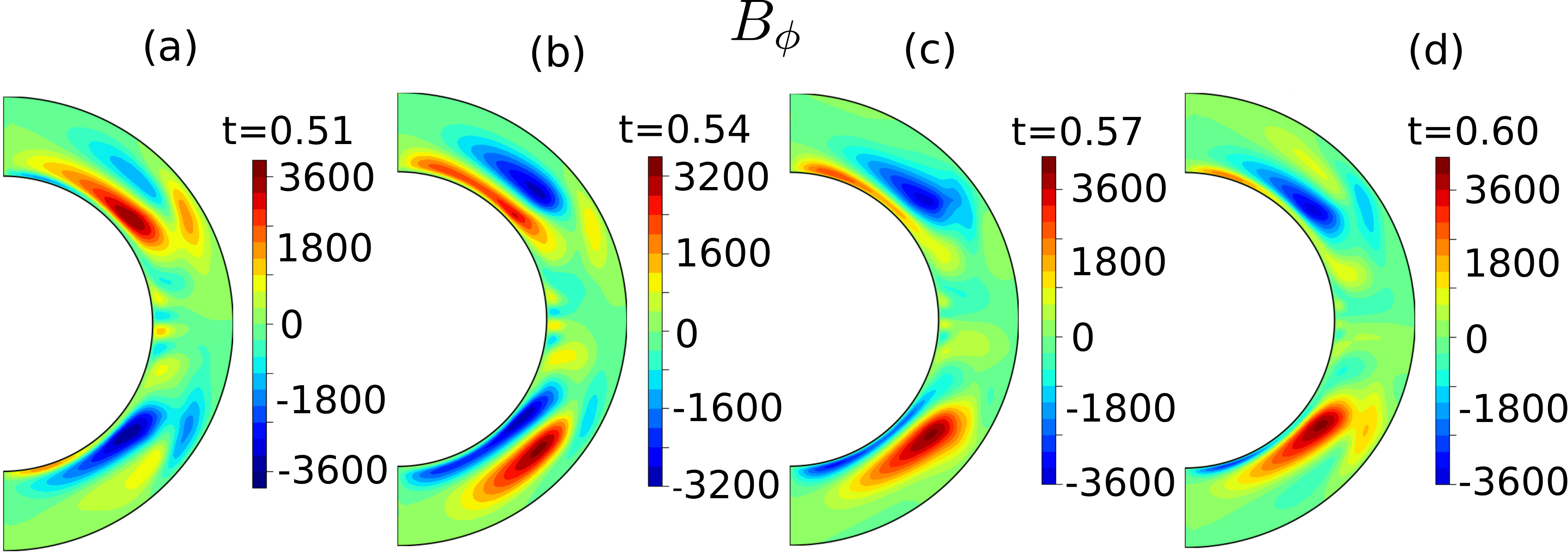}
\caption{Snapshots of the longitudinally averaged toroidal magnetic field ($B_\phi$) at different stages during half a solar magnetic cycle. Subfigures show the polarity reversals of the toroidal field. Here, the magnetic field is in a non-dimensional unit, a value of 1 corresponds to $20$ Gauss.}
\label{fig:bp_avg}
\end{figure*}

\begin{figure*}[htbp]
\centering
\includegraphics[scale=0.45]{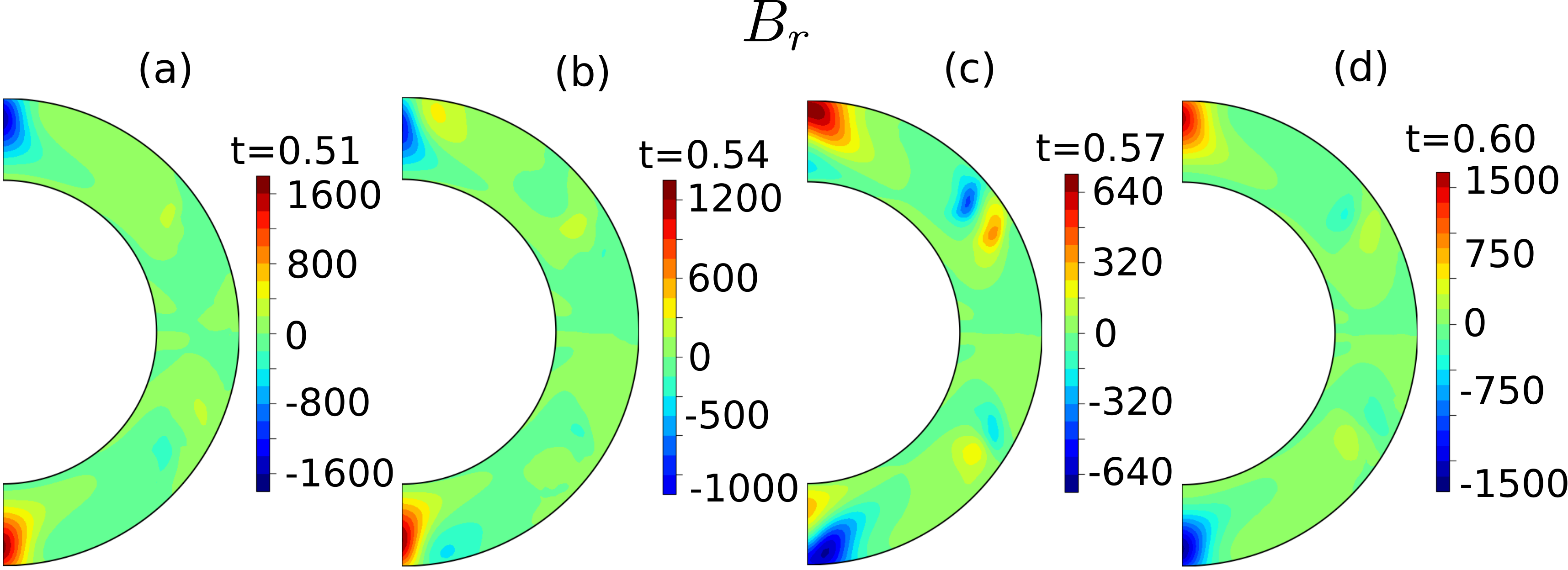}
\caption{Snapshots of the longitudinally averaged non-dimensional radial magnetic field ($B_r$) at different stages during half a solar magnetic cycle. Subfigures show the polarity reversals of the polar field. Here, a unit magnetic field corresponds to $20$ Gauss.}
\label{fig:br_avg}
\end{figure*}

Fig.~\ref{fig:bp_avg} shows snapshots of the longitudinally averaged toroidal magnetic field during half a magnetic cycle (the time duration between two consecutive polarity reversals). Initially, the toroidal field has a positive (resp. negative) polarity in the northern (resp. southern) hemisphere, as seen on the first panel [Fig.~\ref{fig:bp_avg}(a)]. In Fig.~\ref{fig:bp_avg} (b), a new negative polarity field starts to be amplified in the northern hemisphere and gets transported through advection by the meridional flow and diffusion inside the convection zone. At a later time, this newly generated toroidal field occupies the main part of the convection zone [Figs.~\ref{fig:bp_avg} (c) and (d)] and will serve as a seed for the emergence of new BMRs of opposite polarity. In Fig.~\ref{fig:br_avg}, we illustrate the snapshots of the longitudinally averaged radial magnetic field at the same times as in Fig.~\ref{fig:bp_avg}. We again clearly see the reversal of the polar field in time. At $t=0.51$, the dominant toroidal field is positive in the northern hemisphere [Fig.~\ref{fig:bp_avg}(a)], creating BMRs with a positive trailing polarity at the surface. This positive polarity is then advected towards the poles to reverse the negative polar field, as seen in Fig.~\ref{fig:br_avg}(b). When the negative toroidal field starts to be dominant [Fig.~\ref{fig:bp_avg}(c)], it is now BMRs with a negative trailing spots which will emerge, as seen in Fig.~\ref{fig:br_avg}(c). It will then again start to cancel the positive polar field [Fig.~\ref{fig:br_avg}(d)] and finally reverse it to start a new cycle. We note here that the typical values of the poloidal field at the poles is very high (of the order of several kiloGauss) compared to real solar observations. This was already observed in a similar model in \cite{Miesch:ASR2016}. This is one of the shortcomings of our model. We know however that this amplitude will be strongly related to our buoyancy algorithm and for example reducing the number of emerging spots or the frequency of emergence will reduce the polar field by the same amount. A full parametric study and a calibration on the real Sun will be the subject of a future article.

\subsection{Role of the BMRs in the dynamo cycle}

\begin{figure}[htbp]
\centering
\includegraphics[scale=0.65]{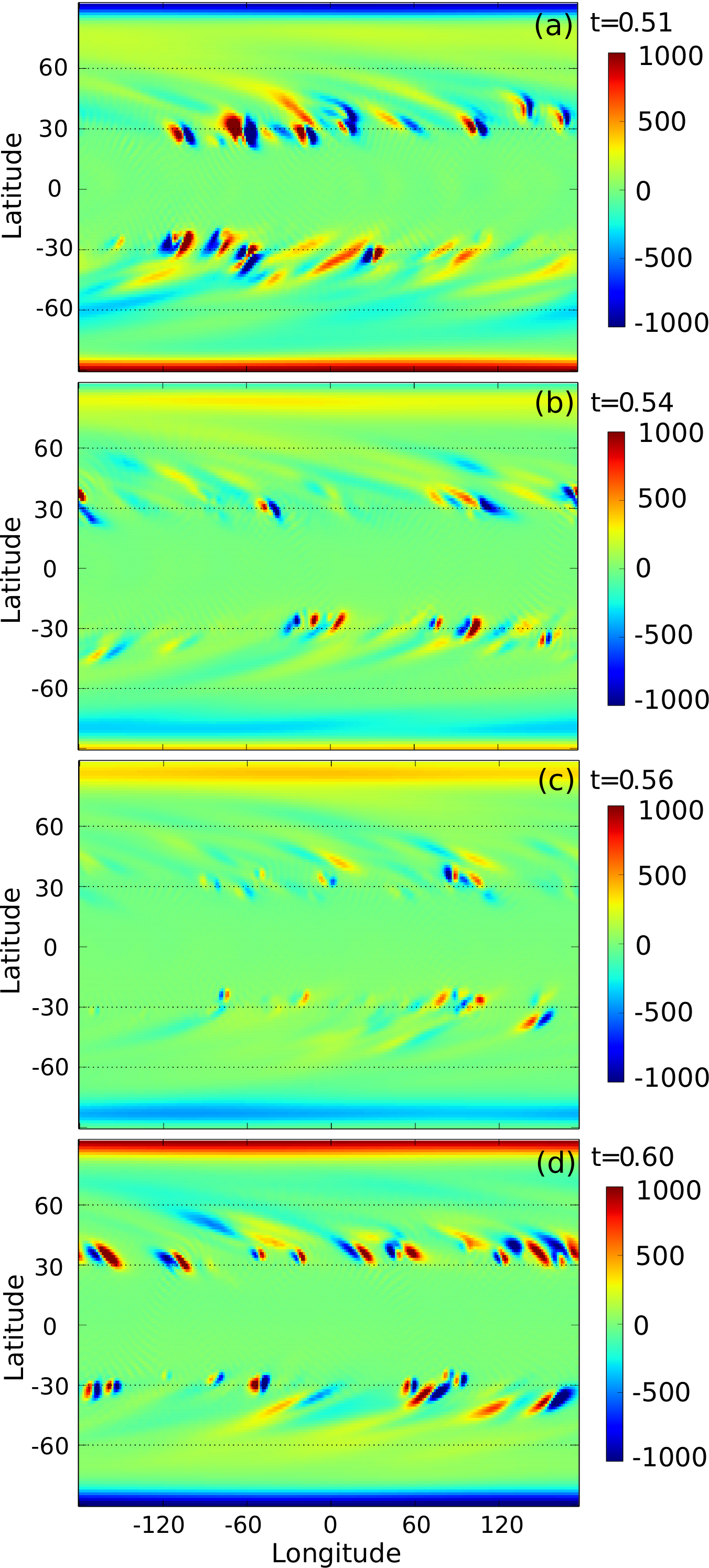}
\caption{Snapshots of the non-dimensional radial magnetic field ($B_r$) at the surface showing (a) the large-scale polar field along with tilted BMRs, (b) reversal of the polar field, (c) sunspot minimum, and (d) the polar field and the BMRs after the polarity reversals. Here, a unit magnetic field corresponds to $20$ Gauss.}
\label{fig:bmr_rev}
\end{figure}

The buoyancy acting on the toroidal field near the base of the convection zone results in the production of multiple tilted BMRs at the surface. The distribution of radial magnetic field at the surface as a function of time is shown in Fig.~\ref{fig:bmr_rev}. The horizontal axes represent the longitude and the vertical axes the latitude. In this figure, we can clearly see the process through which the polar field reverses due to the net flux of BMRs. Indeed, a new set of BMRs gets generated randomly at different longitudes and latitudes every 4.5 months so that we have enough magnetic flux available at the surface in order to reverse the poloidal field. The net magnetic flux of the BMRs at the surface gets advected towards the poles due to the meridional flow and magnetic diffusion, which leads to the polarity reversals of the previous poloidal field at the poles (see Fig.~\ref{fig:bmr_rev}). Subsequently, the shearing of poloidal field due to the differential rotation will reproduce the toroidal field at the base of the convection zone, completing the whole magnetic cycle in our model. 

As expected, the polarities of the produced BMRs change along with the polarity of the toroidal field near the base of the convection zone. As the polarity of the toroidal field reverses in the two hemispheres, the polarities of the generated BMRs also reverse. In Fig.~\ref{fig:bmr_rev}(a), we see the BMRs generated by a toroidal field of positive polarity in the northern hemisphere (positive trailing spot), whereas Fig.~\ref{fig:bmr_rev}(d) demonstrates that the BMRs produced by a toroidal field of negative polarity now have a negative trailing spot. We see here the BL mechanism at play, i.e., the ability of BMRs to be advected towards the poles to reverse the field of the previous cycle. In this model, BMRs thus play a crucial role in the cyclic behaviour of the magnetic field. Note however that in our model, only few bipolar spots appear at latitudes lower than $\pm 30 $ degrees. This is another shortcoming of our model which is being studied at the moment. This can be adjusted by modifying the meridional circulation profile and speed at the base of the convection zone. Indeed, in this current model, the new toroidal field gets generated at high latitudes and transported to the surface to produce spots before getting significantly advected towards the equator by the meridional flow (see Fig.~\ref{fig:bp_avg}). Therefore, the strong toroidal field does not really reach low latitudes and thus spots are not produced at those latitudes, contrary to what is observed in the Sun. By changing the meridional flow profile however, a stronger advection of the toroidal field towards the equator can be achieved and more spots can emerge at low latitudes. At this stage though and as stated above, we are not trying to calibrate the model to real solar observations, this will be the subject of future work. 

\begin{figure}[htbp]
\centering
\includegraphics[scale=1.1]{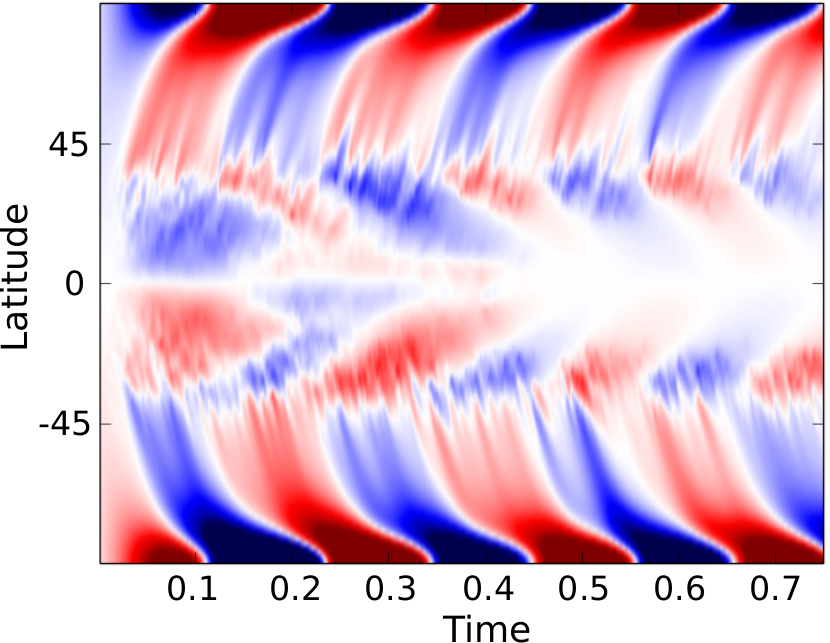}
\caption{Butterfly diagram for the mean radial magnetic field ($Br$) at the surface showing field reversals. The horizontal axis represents the magnetic diffusion time.}
\label{fig:butterfly}
\end{figure}

To get an idea about the length of the magnetic cycle, we plot the longitudinally-averaged surface radial magnetic field as a function of time and latitude, known as the butterfly diagram. This is illustrated in Fig.~\ref{fig:butterfly}. We note that since the BMRs are tilted, the longitudinally averaged flux at mid-latitudes is non-zero, the tilt is thus crucial for the BL mechanism to work since a net flux needs to be advected towards the poles. The advection of this net flux is clearly visible in Fig.~\ref{fig:butterfly} above 40 degrees and all the way to the poles where the previous polarity eventually reverses. In this case, the average time for the poloidal field reversal is almost $8$ years, which makes the length of the complete magnetic cycle equal to $16$ years. In this model, the length of the cycle is highly sensitive to the strength of the meridional circulation; it decreases for stronger meridional flow. We have not tried to calibrate our model on the Sun yet, but we know that modifying the meridional flow amplitude by some fraction, keeping in the range of values observed on the Sun, will help us get closer to an 11-yr cycle.

\section{Corona and solar wind}
\label{sec:corona_wind}  

\subsection{Magnetic topology}  
\label{sec:cor_top}  

\begin{figure}[htbp]
\centering
\includegraphics[width=0.30\textwidth]{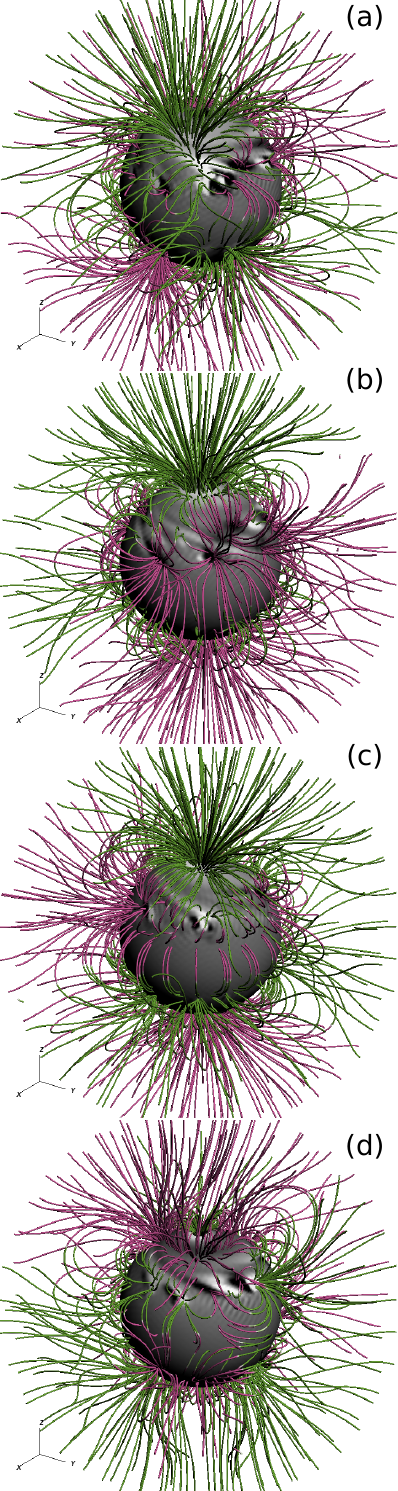}
\caption{Magnetic field lines at the solar surface at the rise phase of the fifth cycle, at the reversal of the polar field, at sunspot minimum, and immediately after the polarity reversal (as in the previous figures). The surface $B_r$ is rendered in gray scale (white for positive and black for negative polarities). The lines are magnetic field lines extrapolated from the surface using PFSS, with source-surface radius at $2.5\ \mathrm{R_\odot}$. Green and violet colours correspond to positive and negative $B_r$.}
\label{fig:mag_field_lines}
\end{figure}

\begin{figure}[htbp]
\centering
\includegraphics[width=0.45\textwidth]{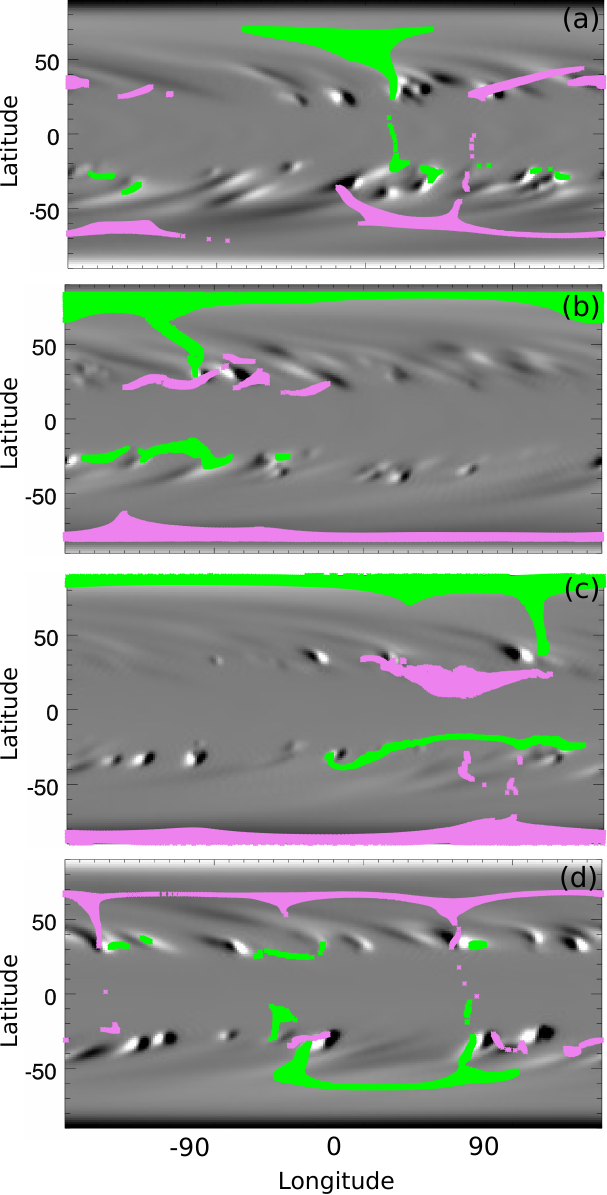}
\caption{Maps of the source regions of coronal holes at the surface. The colour scheme and the instants represented are the same as in Fig.~\ref{fig:mag_field_lines}.}
\label{fig:mag_coronal_holes}
\end{figure}

\begin{figure}[htbp]
\centering
\includegraphics[width=0.45\textwidth]{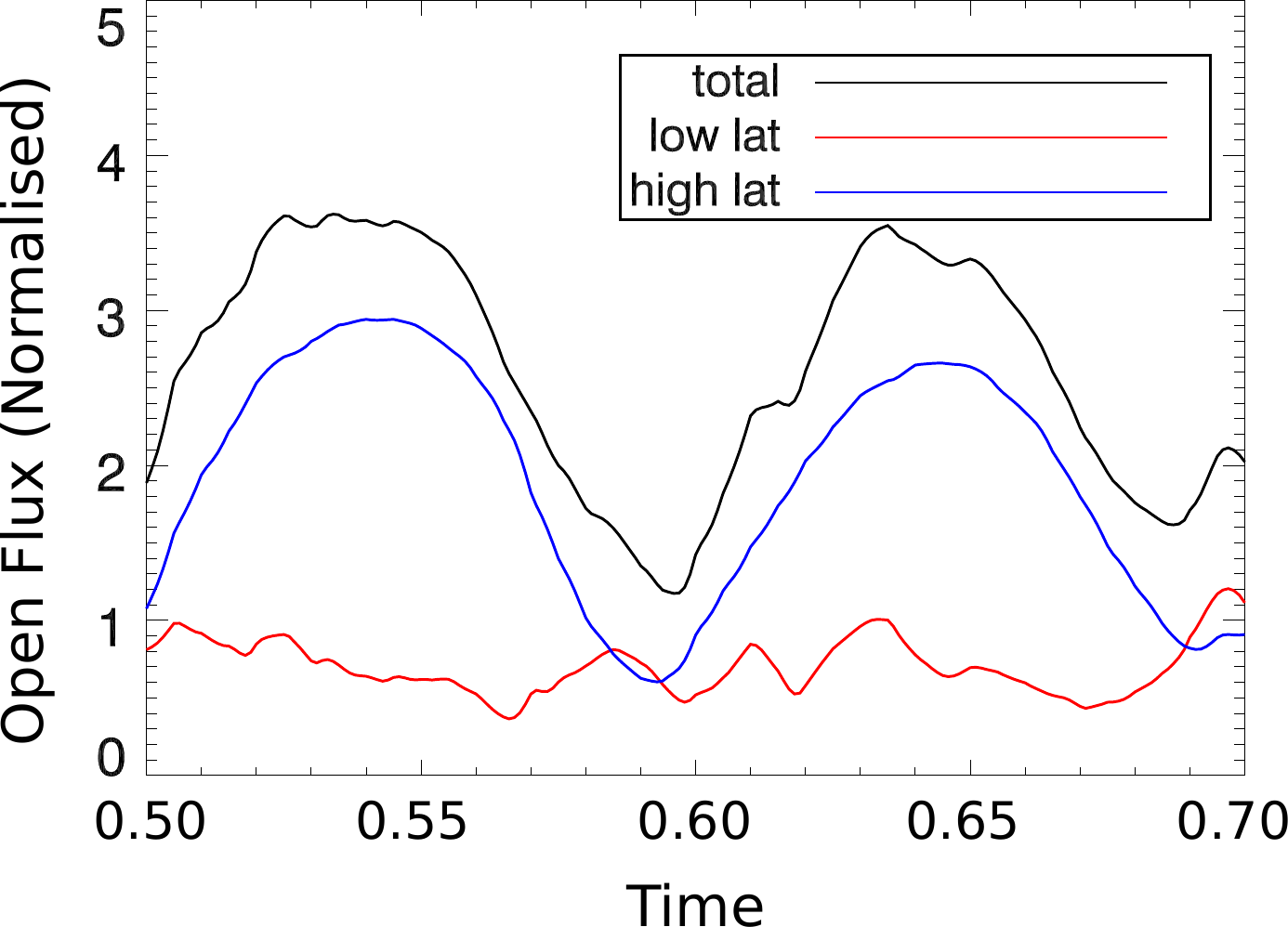}
\caption{Open magnetic flux as function of time for the two half magnetic cycle. The black line shows the total open flux measured at the source-surface (at $2.5\ R_\odot$), while the blue and red lines represent high and low latitude open flux (beyond or within $\pm 45\degree$.) The high latitude open flux anti-correlates with the sunspot cycle phase, and dominates the total flux throughout most of the cycle.}
\label{fig:mag_open_flux}
\end{figure}

The formation and evolution of BMRs and the slower variation of the large-scale field both impact the large-scale topology of the magnetic field of the solar corona. In order to analyze these effects, we extrapolated the surface magnetic field computed by the dynamo model to the solar corona by applying the Potential-Field Source-Surface (PFSS) method. We set a constant source-surface radius $R_{SS} = 2.5\ R_\odot$ for all the extrapolations and follow the extrapolation methods described by Wang and Sheeley~\cite{Wang:APJ1992}. 

Fig.~\ref{fig:mag_field_lines} shows a sequence of snapshots of the coronal magnetic field at times $t = 0.51,\ 0.54,\ 0.56,\ 0.60$  (as in the previous figures). The radial component of the magnetic field at the surface is represented in gray scale (between $\pm 6 \times 10^3$ \ Gauss), with black and white representing respectively the negative and positive polarity. The field line colours indicate that the radial component of the coronal magnetic field is either positive (green) or negative (violet). During the rise phase of the cycle (first panel in the figure), emerging BMRs perturb the quasi-dipolar configuration of the background coronal field, composed mostly of large trans-equatorial magnetic loop systems (streamers) and polar coronal holes which open up with height and end up filling up all the space above (expanding to all latitudes and longitudes at $2.5\ R_\odot$). The newly emerged magnetic flux systems start by gently bending this quasi-bimodal distribution, making streamers incur into higher latitudes and coronal holes extend towards lower latitudes. This continuous distribution of topological elements breaks up later on as the cycle proceeds, with new streamers appearing frequently at mid latitudes at sunspot maximum (second panel) and other closed-field structures (such as pseudo-streamers) even appearing at the polar regions. After the global polarity reversal, these high-latitude loop systems gently fade away, letting the corona relax again toward a simpler configuration until the following minimum.

Some BMRs produce a strong imprint on the overlying coronal fields even much after their emergence at the surface. Shearing by differential rotation elongates the emerged flux in azimuth until they get diffused away. Strong enough BMRs produce long polarity inversion lines that last long enough to sustain loop arcade systems that extend above them (e.g., in the two last panels, on the northern hemisphere, left side of the image). Overall, there seems to be a rather large fraction of closed to open magnetic flux between the solar surface and the source-surface, with open flux regions being rooted on small portions of the surface.
This is more easily visible in Fig. \ref{fig:mag_coronal_holes}, which shows maps of the regions of the surface that are magnetically connected to the coronal holes that form above them. The colour scheme is the same as in Fig. \ref{fig:mag_field_lines}. In comparison to the real Sun, the total extent of sources of open field is small, especially the polar regions at solar minimum. In fact, the emerging regions hold very intense and concentrated magnetic fluxes. Fig. \ref{fig:mag_open_flux} shows the variation of the total unsigned open magnetic flux (measured at the source surface) as a function of time for the whole duration of the dynamo simulation. The high and low latitude components of the open flux (above and below $\pm 45\degree$ in latitude) are also shown as blue and red lines, respectively.
The total open flux is dominated by its high-latitude component through most of the solar cycle, the only exceptions occurring at sunspot maximum,  close to polarity reversal. As expected, the high-latitude component is clearly anti-correlated with the sunspot cycle, while the low-latitude component shows no such clear correlation, being only marginally stronger at sunspot maximum (see Wang and Sheeley~\cite{Wang:APJ2009}).

\subsection{Solar wind}
\label{sec:wind}

\begin{figure}[htbp]
\centering
\includegraphics[width=0.45\textwidth]{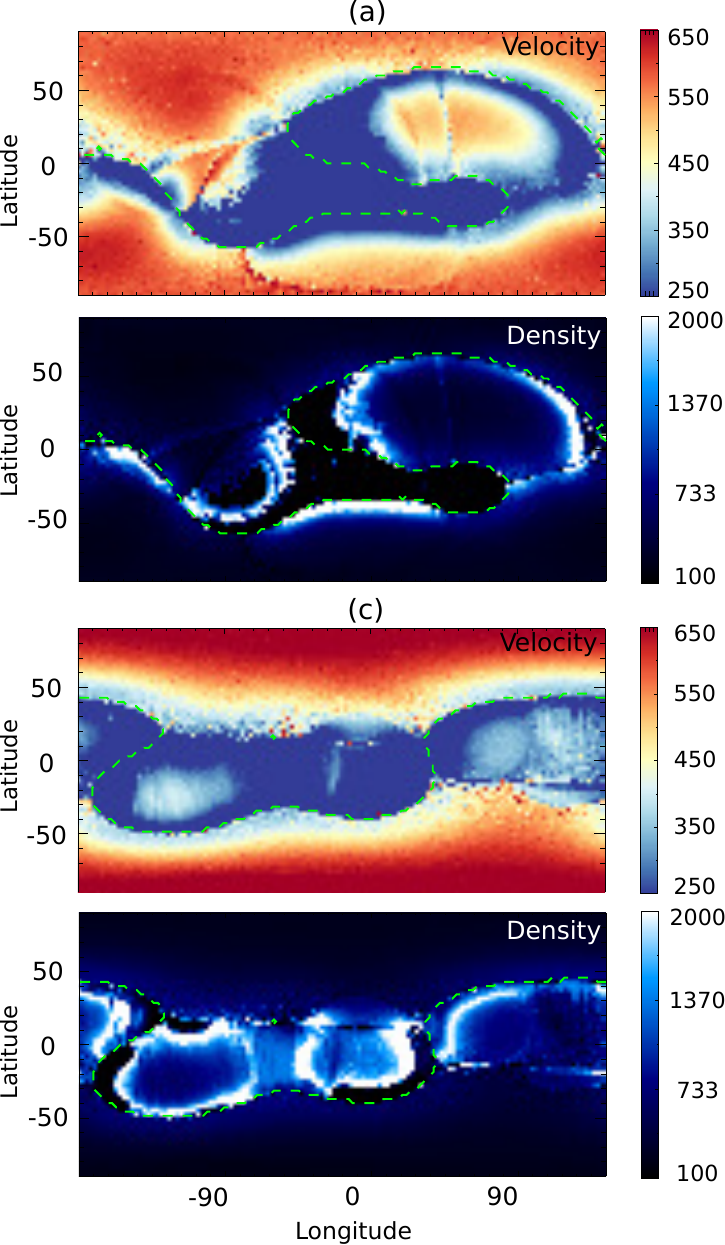} 
\caption{Maps of the velocity (in $km/s$) and density (in $cm^{-3}$) of the solar wind at $r = 21.5\ R_\odot$ corresponding to panels (a) and (c) in Figs.~\ref{fig:mag_field_lines} and~\ref{fig:mag_coronal_holes}. The green dashed lines indicate the position of the polarity inversion line.}
\label{fig:wind_maps}
\end{figure}

The variations of magnetic field geometry in the corona caused by the evolution of the solar dynamo are expected to perturb the large-scale properties of the solar wind flow. We have used the solar wind model MULTI-VP (Pinto \& Rouillard \cite{pinto_multiple_2017}) to analyze the response of the solar wind to these variations. We initiated the wind model by selecting a large collection of open magnetic flux-tubes from the PFSS extrapolations of the dynamo surface fields (see Sec. \ref{sec:cor_top}). The flux-tubes are seeded uniformly throughout the source-surface with an angular resolution close to $3\degree$ (corresponding to more than $8000$ flux-tubes per single full map). The model computes a solar wind profile from the surface of the Sun up to $r=31 \, R_{\odot}$ for each individual flux-tube, fully taking its geometry into account (field amplitude, expansion and inclination profiles). Finally, the one-dimensional solutions are reassembled into a spherical data-cube spanning the whole solar atmosphere. Figure \ref{fig:wind_maps} shows a series of maps representing the wind speed and density at $r=21.5 \, R_\odot$, at two different instants of the simulation (close to sunspot maximum and minimum). The wind speed is expressed in $km/s$ and density in $cm^{-3}$. The green dashed lines represent the position of the polarity inversion line in the corona above the source-surface, which indicates the shape and location of the heliospheric current sheet (HCS). Panel (a) of the figure shows a highly warped HCS, as can be found in the solar atmosphere for periods close to sunspot maximum. Panel (b) shows a more moderately warped HCS that is trying to relax to a configuration more typical of solar minima. 

We obtain in all the cases a distribution of slow and fast solar wind streams (respectively below and above $\sim 450\ \mathrm{km/s}$) that displays many of the main properties of the actual solar winds. The high-speed wind streams are mostly concentrated at high latitudes, being formed within polar coronal holes (see Fig. \ref{fig:mag_coronal_holes}), with the highest wind speeds (reaching more than $750\ \mathrm{km/s}$) being attained during solar minimum. Some streams of fast wind also appear at low latitudes during the phases of high solar activity. The transition between slow and fast wind streams is for the most rather sharp (typically only a few degrees wide). However, the total angular extent of slow wind seems to be significantly larger than that of the actual solar wind. But that is consistent with the small coronal hole footpoint areas shown in Fig. \ref{fig:mag_coronal_holes} and discussed in Sec. \ref{sec:cor_top}, which imply a prevalence of flux-tubes with very strong expansions between $r=1$ and $2.5\ R_\odot$ and that stretch over large azimuthal extents, implying that a significant number of wind flows are driven through highly inclined sections during their acceleration (see discussion by Pinto et al.~\cite{pinto_flux-tube_2016}). The wind maps also indicate the presence of wind streams with very low speed ($\sim 200\ \mathrm{km/s}$) in these regions with high magnetic expansions and field-line inclinations which are consistent with the very slow wind streams reported by Sanchez-Diaz et al.~\cite{sanchez-diaz_very_2016} (that should disappear at higher heliospheric altitudes). The wind density is anti-correlated with the wind speed (with the spread being higher in the low speed limit than on the high speed one), in agreement with measurements by HELIOS and ULYSSES spacecraft.

\section{Summary and conclusions}
\label{sec:conclusion}

The magnetic cycle of the Sun produces and modulates the solar wind, affecting space weather around Earth. Other than the available observational data, researchers rely on various kinds of numerical models to understand the solar magnetic cycle and related phenomena. The main objective is to forecast the future solar activity based on past observational data combined with numerical models.

In this paper, we have presented a new 3D flux-transport dynamo model to understand the solar magnetic cycle and how it affects the coronal topology, which is related with the generation of the solar wind. We have used a 3D kinematic dynamo model coupled with a 3D flux emergence model. In our model, the toroidal flux at the base of the convection zone gets buoyantly transported to the outer surface to produce tilted bipolar magnetic regions or sunspots. Later, the decay and dispersal of these sunspots generate a resultant poloidal field. The meridional flow then transports the surface magnetic flux towards the poles to reverse the polarities of the previous polar field. The newly generated poloidal field then gets sheared by the differential rotation to reproduce a toroidal field near the base of convection zone, completing the magnetic cycle. An additional nonlinearity introduced by putting lower and upper cutoffs on the emerging toroidal flux results in the saturation of the magnetic energy, producing a self-sustained saturated dynamo. We note that this model is not intended to reproduce the full complexity of the solar magnetic field but only the evolution of the largest scales. Indeed, the model is for now kinematic (no back-reaction on the velocity field), relies on a particular model (the Babcock-Leighton model) where the effect of convection is not taken into account and finally dynamo action and emergence at small-scales are ignored. 

The combination of a large scale dynamo with localized magnetic flux emergence produces complex topological variations on the magnetic field of the corona throughout the whole activity cycle, that also have strong repercussions in the properties of the solar wind. We have analysed these effects altogether by means of PFSS extrapolations of the surface field produced by the dynamo simulations, and of a new solar wind model (MULTI-VP) constrained by the extrapolations. This novel approach lets us determine the connections between the evolution of the dynamo field at the surface of the sun and that of the corona and solar wind in a quick and precise way which can be easily repeated through the duration of one or several solar cycles. The solar wind model provides a complete set of physical parameters such as the wind speed, density, temperature and magnetic field amplitude  as well as derived quantities such as dynamical pressures and phase speeds, unlike more conventional semi-empirical scaling laws relating speed to magnetic geometry. The model also provides a sophisticated description of the wind thermodynamics between the surface of the Sun and the high corona, unlike most global-scale solar wind models, that are computationally heavier and tend to rely on polytropic MHD or similar approximations. Our approach produces more realistic solar wind properties (e.g., mass fluxes) than polytropic fluid models.
MULTI-VP completely excludes the closed-field regions of the corona (e.g., streamers) from the computational domain, and reduce the physical description of the solar wind models to a collection of flows driven along individual flux-tubes. That way, we do not have to deal with cross-field interactions that would impose strong constraints on the integration time-step that are unnecessary for the kind of steady-state solutions we are looking for.
But more importantly, there are positive trade-offs: this strategy lets us have a much more detailed description of the thermodynamics of the model than we would be able to in a full 3D MHD setup, and the angular distribution of the flows is not limited by magnetic diffusion across the field (which makes, e.g., the transition between slow and fast wind flows excessively wide in global MHD models). These points are discussed more thoroughly in \cite{pinto_multiple_2017}.
We have verified that the model reproduces many of the main features of the coronal field at different moments of the cycle. We reported a slight tendency to underestimate the open to closed magnetic flux ratio (and, correspondingly, overestimate the size of streamers). The general properties of the solar wind are generally very well reproduced, albeit with slow wind regions which are more extended than expected (which is consistent with the low open to closed magnetic flux ratio). 

In conclusion, we have produced here a modeling chain which allows us to track the evolution of internal dynamo processes along with subsequent changes in the coronal and wind structures. This whole modeling chain lets us determine many synthetic observables such as white-light images of the corona (which can be compared to coronograph imagery) and time-series of wind parameters at any position of the solar atmosphere (that can be compared with in-situ measurements by spacecraft).
Some non solar-like features are present in the current setup used here, but these can be improved by modifying the parameters within the framework of our model. For example, there is a very significant overlap between two consecutive magnetic cycles. Hence, this model is not fully capable of producing proper solar minima, which in turn affects the solar wind solutions. Secondly, this model shows only few spots emerging at low latitudes, which can however be modified through another choice of meridional flow profile. The polar field strength is too high compared to the Sun, which can be corrected by modifying the parameters of the buoyancy algorithm (acting on the net emerging flux). In the present model, the rise velocity is constant for toroidal fields $B_{\phi}^l < B_{\phi} < B_{\phi}^h$. However, various studies suggest that the buoyancy should depend on the strength of the toroidal field near the tachocline. First calculations however show that taking this dependency into account does not significantly affect the results presented here. These are the main areas where we need to improve the current model to have a better match with solar observations. This work is thus only a proof-of-concept that a model producing well-defined spots self-consistently coming from a buoyant toroidal field does induce a cyclic reversal of the poloidal magnetic field and sustains dynamo action. We moreover show that this kind of model can be coupled with a solar wind model which shows the evolution of the wind speed and density as a result of the dynamics of the dynamo-generated field.

Future work will involve a more systematic comparison of the model predictions and real data in order to improve the model's methodology. Moreover, this model is a promising candidate for applying data assimilation techniques since it strongly relies on surface  and solar wind features which are observed. Moreover, thanks to our simplifying assumptions and numerical techniques, the whole chain is computationally cheap and thus very appealing for data assimilation where several realizations of the model are needed to minimize the mismatch with observations. Appying data assimilation would enable us to forecast the future solar magnetic activity and related phenomena. Finally, a modified version of this model could also be used to study the magnetic activity of other cool stars with an internal structure and velocity profiles distinct from those of the Sun. 

\section*{Author Contributions}
R. Kumar and L. Jouve developed the flux-transport model, performed the simulations, and studied the solar magnetic cycle. R. F. Pinto and A. P. Rouillard carried out corona and solar wind studies. The manuscript writing was a combined effort of all the authors.     

\section*{Funding}
This work was supported by the Indo-French research grant 5004-1 from CEFIPRA, by the EC FP7 project \#606692 (HELCATS), and by the CNES (project SWiFT).

\section*{Acknowledgments}
Numerical simulations were performed on CALMIP supercomputing facility at Universit\'e de Toulouse (Paul Sabatier), France, under computing projects {\em 2016-P16021} and {\em 2017-P1504}.




\end{document}